\documentclass[10pt,conference]{IEEEtran}

\usepackage{booktabs} 
\usepackage{multirow}
\usepackage{subcaption}
\usepackage{graphicx}
\usepackage{amsmath}
\usepackage{listings}
\usepackage{float}
\usepackage{xcolor}
\usepackage{dblfnote}
\usepackage{url}
\usepackage{float}
\usepackage{cite}
\usepackage{colortbl} 
\usepackage{pgf}

\usepackage[linesnumbered,ruled,vlined]{algorithm2e}

\definecolor{darkgreen}{rgb}{0, 0.5, 0}
\definecolor{gray}{rgb}{0.5,0.5,0.5}
\definecolor{mauve}{rgb}{0.58,0,0.82}
\newcommand{\Comment}[1]{} 

\newcommand{\draftRev}[1]{{#1}}

\definecolor{codegreen}{rgb}{0,0.6,0}
\definecolor{codered}{rgb}{1,0,0}
\definecolor{codegray}{rgb}{0.5,0.5,0.5}
\definecolor{codepurple}{rgb}{0.58,0,0.82}
\definecolor{backcolour}{rgb}{0.95,0.95,0.92}
\definecolor{lightgray}{gray}{0.9}

\lstdefinestyle{mystyle}{
    commentstyle=\color{codegreen},
    keywordstyle=\color{magenta},
    numberstyle=\small\color{black},
    stringstyle=\color{codepurple},
    basicstyle=\footnotesize\ttfamily,
    breakatwhitespace=false,
    breaklines=true,
    captionpos=b,
    keepspaces=true,
    showspaces=false,
    showstringspaces=false,
    showtabs=false,
    tabsize=2
}

\lstdefinelanguage{diff}{
  morecomment=[f][\color{blue}]{@@},     %
  morecomment=[f][\color{red}]-,         %
  morecomment=[f][\color{codegreen}]+,       %
  morecomment=[f][\color{red}]{---}, %
  morecomment=[f][\color{codegreen}]{+++},
}

\newtheorem{theorem}{Definition}

\usepackage{pgf} 

\newcommand{\comparecell}[2]{%
    \pgfmathparse{#1 > #2 ? 1 : (#1 < #2 ? -1 : 0)}%
    \pgfmathtruncatemacro{\comparison}{\pgfmathresult}
    \ifnum\comparison=1
        \cellcolor{green!70}#1%
    \else
        \ifnum\comparison=-1
            \cellcolor{red!20}#1%
        \else
            #1%
        \fi
    \fi
}

\IEEEtriggeratref{44} 

\begin{document}

\title{State Field Coverage: A Metric for Oracle Quality}

\author{
    \IEEEauthorblockN{Facundo Molina\IEEEauthorrefmark{1}, Nazareno Aguirre\IEEEauthorrefmark{2}\IEEEauthorrefmark{3} and Alessandra Gorla\IEEEauthorrefmark{1}}
    \IEEEauthorblockA{\IEEEauthorrefmark{1}IMDEA Software Institute, Madrid, Spain  
    \\facundo.molina@imdea.org, alessandra.gorla@imdea.org}
    \IEEEauthorblockA{\IEEEauthorrefmark{2}University of Rio Cuarto and CONICET, Rio Cuarto, Argentina
    \\naguirre@dc.exa.unrc.edu.ar}
    \IEEEauthorblockA{\IEEEauthorrefmark{3}Guangdong Technion-Israel Institute of Technology, Shantou, China} 
}

\maketitle

\thispagestyle{plain}
\pagestyle{plain}

\begin{abstract}
  The effectiveness of testing in uncovering software defects depends
  not only on the characteristics of the test inputs and how
  thoroughly they exercise the software, but also on the quality of
  the oracles used to determine whether the software behaves as
  expected.  Therefore, assessing the quality of oracles is
  crucial to improve the overall effectiveness of the testing process.
  Existing metrics have been used for this purpose, but they either
  fail to provide a comprehensive basis for guiding oracle
  improvement, or they are tailored to specific types of oracles, thus
  limiting their generality.

  In this paper, we introduce \emph{state field coverage}, a novel
  metric for assessing oracle quality.  This metric measures the
  proportion of an object's state, as statically defined by its class
  fields, that an oracle may access during test execution.  The main
  intuition of our metric is that oracles with a higher state field
  coverage are more likely to detect faults in the software under
  analysis, as they inspect a \draftRev{larger} portion of the object
  states to determine whether tests pass or not.

  We implement a mechanism to statically compute the state field
  coverage metric. Being statically computed, the metric is efficient
  and provides direct guidance for improving test oracles by
  identifying state fields that remain unexamined. We evaluate state
  field coverage through experiments involving 273 representation
  invariants and 249,027 test assertions. The results show that state
  field coverage is a well-suited metric for assessing oracle quality,
  as it strongly correlates with the oracles’ fault-detection ability,
  measured by mutation score. 
\end{abstract}

\section{Introduction}

Improving the reliability of software systems is among the most challenging problems in software engineering. This problem is strongly related to finding software defects, i.e., identifying software behaviors that diverge from the expected behavior. Software testing is one of the most widely used systematic techniques to identify such software defects, and it demands various \draftRev{complex} tasks \cite{DBLP:books/daglib/0020331}. Firstly, software testing requires crafting (manually or in an assisted manner) test inputs that are able to exercise the software under test (SUT) in realistic and sufficiently varied scenarios. Secondly, to increase the automation in test suite execution and checking, it is crucial to capture the intended behavior of software for the designed test cases through \emph{test oracles}, assertions that attempt to capture the expectations on the execution of test cases as accurately as possible. The problem of producing accurate test oracles, known as the oracle problem~\cite{Barr+2015}, has proved to be both difficult and time-consuming.

Constructing accurate test oracles necessarily depends on the intended software behavior and is largely a manual task. It involves the software developers directly, who can greatly benefit from support in oracle construction, particularly via mechanisms to assess oracle quality. Indeed, oracles can be inaccurate either by misrepresenting the developer's intent or by being too weak, i.e., approximating the intended behavior in a way that fails to reveal many faults. An effective technique to assess oracle quality can direct developers to inaccuracies and other limitations, helping them produce stricter and more accurate oracles with an improved ability to detect defects. 

Various approaches have been proposed to evaluate oracle quality. Among these, mutation testing~\cite{DBLP:journals/ac/PapadakisK00TH19,DBLP:books/daglib/0020331}, a well-established technique for assessing test suite effectiveness, has been widely adopted due to its ability to approximate the fault-detection capability of tests and oracles. Additionally, more direct metrics have been introduced, such as \emph{checked coverage}~\cite{DBLP:conf/icst/SchulerZ11}, which measures how thoroughly test assertions cover SUT statements that influence their outcomes, and the search-based oracle deficiency detection put forward in \cite{DBLP:conf/issta/JahangirovaCHT16}, which quantifies assert statement quality by identifying false positives (correct executions incorrectly flagged as erroneous) and false negatives (undetected faults) that the oracles lead to.

Despite these advances in oracle assessment, existing approaches suffer from various limitations. These techniques typically rely on dynamic analyses that are computationally expensive. In particular, mutation analysis requires repeated test executions to identify killed and surviving mutants; checked coverage depends on dynamic slicing to trace SUT statements affecting oracles; and search-based oracle deficiency detection combines evolutionary computation~\cite{DBLP:books/daglib/0019083} with mutation analysis, both computationally expensive tasks. Furthermore, while these techniques provide accurate oracle quality metrics, their feedback for improving oracles remains indirect \draftRev{(e.g., surviving mutants or unassessed SUT statements, whose translation into oracle improvements is non-trivial)}.  

In this paper, we introduce \emph{state field coverage}, a novel metric for oracle assessment. Our approach evaluates oracle quality by measuring the proportion of the SUT’s state definition (i.e., class fields, in the context of object-oriented code) referenced by the oracle. A field is considered ``covered'' if it is directly or indirectly referred to in oracle expressions. As opposed to existing techniques to assess oracle quality, we present an approach that computes our metric statically, thus avoiding costly dynamic analyses. Moreover, our approach provides actionable feedback by explicitly identifying uncovered state fields, guiding oracle enhancements. \draftRev{Also, while other metrics (e.g., mutation testing) do yield explicit feedback, they do not easily indicate oracle improvements. For example, surviving mutants indicate undetected faults, but require providing additional test data or crafting new tests to kill these mutants~\cite{mutant-killing-2023}, which are typically non-trivial to provide. In contrast, state field coverage identifies specific state variables omitted from oracles, directing the improvement process.} 

Our experiments, comprising oracles originating from 273 representation invariants and 249,027 test assertions, show that state field coverage is an effective metric for assessing oracle quality, as it strongly correlates with fault-detection ability measured by mutation score.

\section{Related work}

This section discusses some of the established approaches to assess the quality of oracles,
and other related works that are relevant to our proposal.




\subsection{Oracle Quality Assessment}

\subsubsection{Mutation-based Oracle Assessment}

Mutation testing~\cite{DBLP:journals/ac/PapadakisK00TH19,DBLP:books/daglib/0020331} is a widely used technique to assess the quality of test suites, in terms of the ability of test suites to detect (artificial) faults in the SUT. More precisely, mutation testing generates mutants of the SUT by injecting artificial faults in the code, and then measures the proportion of mutants that are detected (or killed) by the test suite, i.e., that make at least one test case fail. The proportion of mutants that are killed, known as the \emph{mutation score}, is known to correlate with real fault detection better than other traditional testing metrics~\cite{DBLP:conf/sigsoft/JustJIEHF14, DBLP:conf/icse/PetrovicIFJ21}. This score has also been effectively used as a metric to assess the quality of oracles in a variety of techniques for test oracle automation~\cite{DBLP:conf/issta/FraserZ10, Molina+2021,Molina+2022, BlasiGEPC2021, gassert2020, DBLP:conf/issre/GargDMCAPT23, DBLP:conf/issta/AlonsoSR23, DBLP:journals/tosem/MolinaGd25, DBLP:journals/tse/HayetSd25, DBLP:conf/icse/HossainD25} and oracle assessment studies~\cite{DBLP:conf/issta/JahangirovaCHT16, DBLP:conf/sigsoft/HossainFDEV23, DBLP:conf/sigsoft/ZhangM15,DBLP:conf/icst/SchulerZ11}. These works follow an approach similar to the use of mutation for test suite assessment: given the SUT and an oracle (e.g., an assertion) for it, the quality of the oracle is measured by its ability to \emph{kill} mutants, i.e., to identify mutants through violations to the oracle, e.g., on generated or provided test suites. Through this analysis, one can measure the \emph{strength} of the oracles in terms of their ability to detect faults.

Mutation-based oracle analysis can be effective in revealing potential oracle weaknesses, by identifying faults that the oracles are not able to detect. However, it also has some limitations. Firstly, the result of the mutation analysis points to the mutants that are not detected by the oracle, but exploiting this feedback to improve the oracle itself (i.e., to recognize the relationship between mutants and improvements to the oracle) is non-trivial. Secondly, oracle assessment based on mutation usually combines test generation with mutation analysis, both computationally expensive tasks, notably the latter. 

\subsubsection{Coverage-based Approaches}
 
Checked coverage~\cite{DBLP:conf/icst/SchulerZ11} 
is a metric that focuses on the evaluation of the quality of test assertions. The checked coverage metric essentially works by analyzing the code features that affect the expressions involved in test oracles. To \draftRev{calculate} checked coverage, a dynamic backward slice of the test oracles is computed, which determines the statements that contribute to the checked expressions. The percentage of statements that contribute to the expressions checked in the test oracles, in relation to the total number of statements of the \draftRev{SUT}, constitutes the checked coverage. In other words, it attempts to measure the proportion of sentences of the \draftRev{SUT} whose effect is being directly or indirectly checked by the test assertions.

Checked coverage has proved to be a good indicator of test oracle quality, correlated with the ability of the oracles to detect faults and even more sensitive than mutation score~\cite{DBLP:conf/icst/SchulerZ11}. However, checked coverage concentrates on the evaluation of test assertions, and is not easily adaptable to support other more general types of oracles, such as contract assertions (e.g., pre and postconditions)~\cite{Meyer1997, DBLP:journals/computer/Meyer92}. Consider, for instance, a postcondition. Since the assertion is local to a method, it seems more reasonable to measure checked coverage with respect to sentences only of the method itself; the generality of such assertions together with their locality to specific methods would typically lead to high checked coverage values. Additionally, \draftRev{computing checked coverage is expensive}, demanding from several hours to days for large projects~\cite{DBLP:conf/ast/Koitz-HristovSW22}, since it requires the computation of dynamic slices \cite{DBLP:journals/jss/KorelL90}.

State coverage~\cite{state-coverage} is another approach based on statement coverage. 
As with checked coverage, state coverage also relies on 
program slicing. It measures the quality of checks (e.g., 
test assertions) by considering all output defining
statements (ODS), i.e., statements that define an output variable.
State coverage is computed by counting the number of 
ODS \draftRev{present in} the dynamic slices of a given assertion, 
divided by the total number of ODS.
Being based on ODS, this technique depends 
on the given test inputs (as for different inputs, 
different statements may be output defining).
A positive aspect of the approach is that, in principle,
it can provide ODS that are not being checked
by the assertions, which may be useful for 
users to further improve the oracles.
However, given the very limited evaluation that \draftRev{exists} for this 
technique, just a small experiment with a proof of 
concept implementation in a short paper~\cite{state-coverage}, there is no significant evidence to 
support \draftRev{its usefulness}, or reveal deeper insights or limitations.

A second implementation of state coverage 
has also been proposed~\cite{DBLP:conf/sofsem/VanoverbergheHTP12},
implementing an extension 
in which the state coverage is computed 
as the ratio of state updates that are read by 
assertions with respect to the total number of state updates. 
Every code location in the source code in which 
an update is performed is considered a state update.
To compute state coverage, a test suite is executed 
and monitored using a process that collects 
the set of state updates (writes), \draftRev{and the subset of updates read in the test assertions (reads); state coverage is the ratio of reads over the number or writes.} 


\subsubsection{Oracle Deficiencies}

The deficiencies of an oracle can be determined both 
qualitatively and quantitatively. 
For instance, OraclePolish~\cite{oraclepolish2014} is a dynamic 
technique that qualitatively assesses the quality of test assertions 
by analyzing how they interact with the inputs in a 
specific test. The technique \draftRev{focuses} on detecting brittle assertions, i.e., 
assertions that check values of uncontrolled inputs 
(e.g, inputs declared outside the tests), 
and unused \draftRev{test inputs}, i.e., 
inputs \draftRev{in the tests} that are not checked in the assertions. 
Similarly to the checked coverage metric, OraclePolish 
is also tailored for test assertions, and it is not easily adaptable
to other types of oracles.

The quality of an oracle can be quantitatively assessed by identifying more concrete oracle deficiencies: false positives and false negatives~\cite{DBLP:conf/issta/JahangirovaCHT16}. A false positive is a correct and expected program state for which the oracle fails, i.e., a false alarm; a false negative, on the other hand, is an incorrect and unexpected program state for which the oracle is true, i.e., a missed fault. OASIs~\cite{DBLP:conf/issta/JahangirovaCHT16} is a tool for automatically assessing the quality of oracles, by computing the above oracle deficiencies. It has been used as a metric for oracle quality~\cite{Molina+2021} and to guide the manual as well as the automated improvement of oracles~\cite{DBLP:conf/issta/JahangirovaCHT16,DBLP:journals/tse/JahangirovaCHT21,gassert2020}. OASIs searches for false positives and false negatives using evolutionary computation. False positives are reported as test cases that falsify the oracle when they should not. False negatives, on the other hand, are calculated as mutations that are not detected by the oracle, and thus are based on mutation analysis. Oracle deficiencies provide a more direct input to the improvement of oracles. Their computation is however expensive, and is designed for specific oracles, expressed as assert statements in the code.

In general, the dynamic nature of the existing metrics for oracle quality makes them computationally expensive, and the use of their corresponding results as inputs for oracle improvement may demand subsequent time-consuming analyses. Some of the approaches, e.g., checked coverage and oracle deficiency identification, concentrate on specific kinds of oracles. \draftRev{\emph{State field coverage}, the novel metric for oracle quality that we introduce in Section~\ref{sec:object-state-coverage},} aims to overcome these limitations. Our metric can be efficiently computed, provides an output that more directly leads to oracle improvement, 
and is applicable to different kinds of oracles, including test assertions and contract specifications such as operational class invariants. 

\subsection{Automated Test Oracle Generation}

Most of the metrics for oracle
quality assessment introduced in the previous section
have been used either to
evaluate the quality of automatically inferred
oracles, or to guide an oracle inference process. 
Notably, mutation testing has been the most 
widely used approach 
to assess the quality of automatically derived 
oracles~\cite{Molina+2022, Blassi+2018,BlasiGEPC2021,gassert2020,
memoria2024, mrscout2024, genmorph2024}.
Typically, these oracle generation approaches observe some artifact 
of the SUT, such as
code comments or software executions,
and infer oracles from these artifacts.
Then, mutation analysis is used to assess the 
quality of the inferred oracles by measuring 
their ability to detect artificial faults (mutants).
Moreover, mutation analysis has also been 
used to guide the oracle inference process by trying to maximize the mutation score~\cite{Molina+2021, Molina+2022, DBLP:conf/issta/FraserZ10} or 
by prioritizing the mutants used in the analysis~\cite{DBLP:conf/issre/GargDMCAPT23}, 
leading to more efficient processes and more precise oracles.


More recently, oracle deficiencies have 
also been utilized
as part of oracle inference processes.
In particular, the OASIs tool has been used
as a core component in a technique that implements an evolutionary approach that 
tries to infer assertions minimizing the 
number of false positives and false negatives~\cite{gassert2020}.
Though effective, computing oracle deficiencies using 
dynamic analysis, as in the case of OASIs, is computationally expensive~\cite{gassert2020}. 

Our state field coverage metric 
may also be used to evaluate the quality of automatically 
generated oracles, as well as to guide oracle inference processes by optimizing 
the state field coverage. In fact, as it is possible to compute our metric statically,
it is expected to be more efficient than the existing metrics,
and thus more suitable for guiding the oracle inference process. 
We plan to explore this application in future work.

\section{The State Field Coverage Metric}
\label{sec:object-state-coverage}

In this section, we formally introduce \emph{state field coverage}, our proposed metric to assess oracle quality. This metric is based on the idea of analyzing to what extent a given oracle predicates over the state of the software under analysis. The intuition here is rather straightforward: the more an oracle examines the software state, the better the oracle is. Instead of assessing state field coverage in a dynamic fashion, our approach concentrates on how the state is statically defined. Indeed, from the state definition of the software under analysis, e.g., in an object oriented setting, the definitions of fields in the classes that compose the software, we build a graph-like summary that we call the \emph{type graph}, and statically analyze the proportion of the type graph that is involved in the oracle, i.e., the direct and indirect fields present in the oracle expressions. Below we formally define these concepts, and show how our metric is computed.

Let us first assume that the software under analysis is organized as a collection $C_{1}, \dots , C_{n}$ of classes, where $C_{1}$ is a distinguished \draftRev{root} class. Each class $C_{i}$ defines a set of fields, whose types are among $C_{1}, \dots , C_{n}$ and primitive datatypes. From a given class $C$, we can compute the set $F_{c}$ of reachable fields, which includes all the fields in $C$ plus all the reachable fields of the classes $C_{i}$ for which there is a field in $C$ of type $C_{i}$. A field $f \in F_{c}$ is \emph{iterable} if its type is a collection (e.g., arrays, sets, etc), and \emph{non-iterable} otherwise. Moreover, recursive fields (fields from a class to itself), and other fields present in classes that contain at least one recursive field 
(e.g, value fields in nodes), 
are also considered iterable, as they enable iteration over objects of \draftRev{linked structures}. The set of iterable fields is denoted by $I_{c}$. 

\subsection{Type Graph}

\begin{figure}[t]
\begin{minipage}{.45\textwidth}
\centering
\footnotesize
\begin{lstlisting}[language=Java,linewidth={\linewidth},frame=tb]
public class LinkedList<E> extends AbstractSequentialList<E> implements ... {
  int size = 0;
  Node<E> first, last;
  ...
  private static class Node<E> {
    E item;
    Node<E> next, prev;
    ...
\end{lstlisting}
\caption{LinkedList class from the java.util package.}
\label{fig:linked-list-java-util}
\end{minipage}
\hfill
\begin{minipage}{.40\textwidth}
\centering
\includegraphics[width=\textwidth]{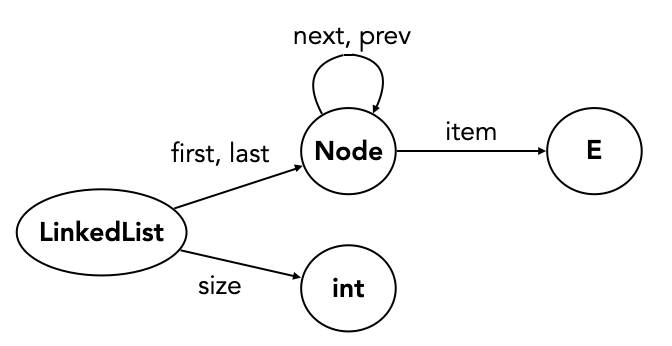}
\caption{Type graph of the LinkedList class.}
\label{fig:linked-list-type-graph}
\end{minipage}
\end{figure}


A \emph{type graph}, introduced in~\cite{Molina+2019}, is an abstract representation of a class that captures the relationships between the types of all the reachable fields of a software under analysis.

\begin{theorem}
\label{def:type-graph}
Given a class $C$, its \emph{type graph} $G_{c}$ is defined as the structure $(V_{c}, E_{c})$, where $V_{C}$ is a set of nodes representing types ($C$ and all the types of the fields reachable from $C$), and $E_{c}$ is the set of edges representing the reachable fields, i.e., for each field $f$ of type $T$ in a class $C_{i}$, there is an arc in the graph going from the node representing $C_{i}$ to the node representing $T$. 
\end{theorem}


As an example, consider the \texttt{LinkedList} class from the java.util package shown in Figure~\ref{fig:linked-list-java-util}. The class declares three fields: \texttt{first} and \texttt{last} of type \texttt{Node}, and \texttt{size} of type \texttt{int}. Additionally, the inner class \texttt{Node} also declares three fields: \texttt{next} and \texttt{prev} of type \texttt{Node}, and \texttt{item} of the generic type \texttt{E}. Figure~\ref{fig:linked-list-type-graph} shows the type graph for the \texttt{LinkedList} class. This graph contains four nodes, one per each reachable type (\texttt{LinkedList}, \texttt{Node}, \texttt{int} and \texttt{E}). Then, for each of the mentioned fields, there is an arc connecting the nodes representing the corresponding types. For this example, fields \texttt{first}, \texttt{last} and \texttt{size} are non-iterable, while \texttt{next}, \texttt{prev} and \texttt{item} are iterable, since the first two are recursive, and the last belongs to a class with a recursive field. Notice that these fields are considered iterable as they allow one to iterate over the nodes and values of a linked list. In fact, at run time, they can be considered to lead to the definition of sets of elements of type \texttt{Node} and \texttt{E}, respectively, e.g., all the nodes reachable through \texttt{next} (resp. \texttt{prev}) from a given node, or all items obtained from the nodes through field \texttt{item}.

\subsection{State Field Coverage}

To provide a formal definition of the state field metric,
we define the concepts of \emph{coverable and covered labels}, i.e., 
the set of target labels that an oracle \draftRev{may} cover, and the subset of these that the oracle actually covers, respectively. 

\begin{theorem}
  \label{def:coverable-labels}
  Let $C$ be the target class and $G_{c}$ its type graph. 
  We define the set $L_{c}$ of \emph{coverable labels}
  as $E_{c} \cup S$, where $E_{c}$ is the set of edges
  (fields) in the type graph $G_{c}$ and $S$ is the set 
  of special labels, defined as $S = \{ f+ \mid f \in I_{c} \}$,
  computed from the set $I_{c}$ of iterable fields.
\end{theorem}

\draftRev{Basically}, the set of coverable labels $L_{c}$ includes
\draftRev{a} label $f$ for each field $f \in F_{c}$,
and \draftRev{a} special label $f+$ for each iterable field $f \in I_{c}$. 
For the \texttt{LinkedList} example, the set 
of coverable labels will contain the field labels 
$\lbrace first, last, size, next, prev, item \rbrace$,  
and the special labels $\lbrace next+, prev+, item+ \rbrace$ 
corresponding to the iterable fields \texttt{next}, \texttt{prev}
and \texttt{item}.

\begin{theorem}
\label{def:covered-labels}
Let $p$ be a program taking as input 
an object of class $C$. We say that a label 
$l \in L_{c}$, corresponding to field $f$,
is covered by $p$, if $p$ accesses field $f$. A label $l+ \in L_{c}$, corresponding to an iterable field $f$,
is covered by $p$ if $p$ iterates over the elements obtained through $f$. The \emph{covered labels} associated with $p$ is the set of all labels covered by $p$.
\end{theorem}

\draftRev{In the above definition, the covered labels are defined for an arbitrary program $p$. For the context of this paper, the program $p$ will always represent an \emph{oracle}, e.g., the statements that are called, directly or indirectly, within a test assertion.} Given a test oracle, its state field coverage is the proportion
of coverable labels that are actually covered by the oracle.

\begin{theorem}
\label{def:heap-coverage}
Let $\phi$ be an oracle defined for class $C$, i.e., $\phi \colon C \to Bool$.
The \emph{state field coverage} $SFC_{\phi}$ of $\phi$ is defined as
the proportion of coverable labels $L_{c}$ that are 
covered by $\phi$, i.e.,
\begin{equation}
\label{eq:heap-coverage}
SFC_{\phi} = \frac{|L_{\phi}|}{|L_{c}|}
\end{equation}
\end{theorem}

\begin{figure}[t]
  \centering
  \scriptsize
  \begin{subfigure}[t]{.95\columnwidth}
\begin{lstlisting}[language=Java,linewidth={\linewidth},frame=tb]
public boolean isEmpty() {
  return size == 0;
}
\end{lstlisting}
  \caption{Method checking if the list is empty.}
  \end{subfigure}
  \begin{subfigure}[t]{.95\columnwidth}
\begin{lstlisting}[language=Java,linewidth={\linewidth},frame=tb]
public boolean checkSize() {
  if (first == null) return size == 0;
  Set<Node> visited = new java.util.HashSet<>();
  visited.add(first);
  Node<E> current = first;
  while (current != null && visited.add(current)) {
    current = current.next;
  }
  return visited.size() == size;
}
\end{lstlisting}
  \caption{Method checking that the size of the list is correct.}
  \end{subfigure}
  \caption{Two methods 
  over the \texttt{LinkedList} class with 
  different state field coverage.}
  \label{fig:example-invariants-linked-list}
\end{figure}

For example, consider the two methods in
Figure~\ref{fig:example-invariants-linked-list}
defined over \texttt{LinkedList}\draftRev{, and assume that these are called within respective test assertions.} 
The first method checks if the list is empty,
accessing only the \texttt{size} field.
Therefore, 
it only covers the label $size$ out of 
9 coverable labels, which leads to a state 
field coverage of $11.1\%$. 
The second method, on the other hand,
checks that the size of the list is correct,
by accessing the \texttt{size}, \texttt{first} and \texttt{next} fields, and also \emph{iterating}
over the \texttt{next} field.
Thus, it covers 4 labels ($size$, $first$, $next$ and $next+$), achieving an object 
state coverage of $44.4\%$. 

Although state field coverage
is defined for oracles predicating over a single 
class, it can be easily extended to oracles predicating 
over multiple classes. In such cases, 
the coverable labels $L_{c}$ will be the union of the
coverable labels for each class, and the labels $L_{\phi}$ covered by an oracle $\phi$ will be the union 
of the covered labels in each class.

\subsection{Implementation}
\label{sec:osc-impl}

To measure the state field coverage of an oracle, we implement a static analysis approach. Our implementation is for Java. The process takes as input a target class $C$ and the source code of an oracle $\phi_{c}$, and computes the oracle's state field coverage according to Definition~\ref{def:heap-coverage}.  


\subsubsection*{Type Graph \draftRev{Generation}} Given a (root) Java class $C$, we generate the type graph $G_{c} = (V_{c}, E_{c})$ by first initializing the set of nodes $V_{c}$ with a node for $C$, the target class, and then recursively adding edges and nodes for the reachable fields and classes, respectively. 
In our implementation, this process is performed in a depth-first fashion, and the type graph is built using the jgrapht~\cite{jgrapht-site} library.

\subsubsection*{Coverable labels} The set $L_{c}$ of coverable labels is straightforwardly computed from the type graph $G_{c}$. Besides each edge in the graph being a label in $L_{c}$, we consider labels for iterable fields, based on the following two cases: 
\begin{itemize}
  \item For every edge $e = T_{1},T_{2}$ such that $T_{2}$ is a class that implements the \texttt{Iterable} interface, or is an array type, 
  $e$ is deemed iterable, and we add label $l_{e}+$ to $L_{c}$,
  \item For every edge $e = T_{1},T_{2}$ such that $T_{1}$ participates in a loop path (a non-empty graph path starting and ending in the same node) within $G_{c}$, $e$ is deemed iterable, and we add label $l_{e}+$ to $L_{c}$. 
\end{itemize}

\subsubsection*{Covered labels} The labels $L_{\phi}$ that are covered by an oracle $\phi_{c}$ are obtained by parsing the source code of $\phi_{c}$, and identifying the fields accessed by the oracle\draftRev{, i.e., from expressions or methods called within (test) assertions}. More precisely, the following two cases are considered: 
\begin{itemize}
  \item a label $l \in L_{c}$ is considered covered if there exists a statement reachable from $\phi_{c}$'s source code that accesses the field corresponding to $l$,
  \item a special label $l+ \in L_{c}$ is considered covered if there exists a statement in a loop body (e.g., the body of a \texttt{for} or \texttt{while} statement) reachable from $\phi_{c}$'s source, that accesses the field corresponding to $l+$.
\end{itemize}
Oracle source code parsing and analysis is implemented using the JavaParser~\cite{javaparser-site} library. Our current implementation is specific to Java, as it relies on the Java syntax and type system. 
Since our coverage approach assesses how thoroughly the oracles evaluate the SUT’s state definition, its implementation needs to take into account the mechanisms that the programming language provides for data representation. Most programming languages provide means to define custom datatypes, and these lead straightforwardly to notions of type graphs, similar to what we have described above for Java. Although our implementation computes our coverage metric for Java, adapting the process to other programming languages and datatype definition mechanisms is relatively direct. 

\section{Evaluation}

Our evaluation of state field coverage is organized around the following research questions:
\begin{itemize}

\item[\textbf{RQ1}] \emph{Is state field coverage correlated with fault detection?}

\item[\textbf{RQ2}] \emph{Can state field coverage be used for oracle improvement?}

\item[\textbf{RQ3}] \emph{Can oracle improvement based on state field coverage help real bug detection?}

\item[\textbf{RQ4}] \emph{How efficiently can state field coverage be computed?}

\end{itemize}
RQ1 analyzes the correlation between state field coverage and the ability of the oracles to detect faults, measured as the detection of mutants. RQ2 focuses on evaluating how the state field coverage metric can be used to guide the improvement of oracles; we analyze how the ability of test suites to detect artificial faults varies as tests with increasingly larger state field coverage are incorporated, comparing it with fault detection when such tests are randomly added. RQ3 evaluates the impact of state field coverage in detecting real faults, through an experiment similar to the one used for RQ2, but on real regression faults. Finally, RQ4 evaluates the efficiency with which the state field coverage metric can be computed. 

\subsection{Evaluation Subjects}

\begin{figure}[t]
  \centering
  \scriptsize
  \begin{subtable}[t]{0.45\textwidth}
      \centering
      \begin{tabular}{lrr}
        \toprule
        \textbf{Class} & \textbf{\#Properties} & \textbf{\#Rep. Invariants} \\
        \midrule
        SinglyLinkedList & 3 & 7 \\
        SortedList & 4 & 15 \\
        DoublyLinkedList & 3 & 7 \\
        BinaryTree & 3 & 7 \\
        SearchTree & 4 & 15 \\
        RedBlackTree & 5 & 31 \\
        HeapArray & 4 & 15 \\
        BinomialHeap & 5 & 31 \\
        DisjSet & 4 & 15 \\
        FibonacciHeap & 7 & 127 \\
        DAG & 2 & 3 \\
        \midrule
        \textbf{Total} & 44 & 273 \\
        \bottomrule
        \end{tabular}
      \caption{Representation Invariants.}
      \label{tab:subjects-rq1-korat}
  \end{subtable}
  \hfill
  \begin{subtable}[t]{0.45\textwidth}
      \centering
      \begin{tabular}{lrrr}
        \toprule
        \textbf{Project} & \textbf{\#Classes} & \textbf{\#Tests} & \textbf{\#Assertions} \\
        \midrule
Chart & 29 & 6,557 & 27,301 \\
Cli & 6 & 1,008 & 1,347 \\
Closure & 47 & 25,150 & 27,317 \\
Codec & 3 & 2,219 & 5,396 \\
Collections & 6 & 8,160 & 9,764 \\
Compress & 11 & 2,548 & 5,834 \\
Csv & 13 & 840 & 2,748 \\
Gson & 3 & 3,097 & 5,842 \\
JacksonCore & 33 & 1,738 & 10,181 \\
JacksonDatabind & 23 & 6,439 & 21,756 \\
JacksonXml & 6 & 494 & 1,796 \\
Jsoup & 7 & 2,056 & 6,287 \\
JxPath & 9 & 1,152 & 2,052 \\
Lang & 3 & 6,737 & 38,774 \\
Math & 6 & 13,010 & 25,544 \\
Mockito & 4 & 4,064 & 5,364 \\
Time & 6 & 11,998 & 51,724 \\
        \midrule
        \textbf{Total} & 215 & 97,267 & 249,027 \\
        \bottomrule
        \end{tabular}
      \caption{Test Assertions}
      \label{tab:subjects-rq1-defects4j}
  \end{subtable}
  \caption{Distribution of the target representation invariants from Korat (a) 
  and the target test assertions from the 51 project versions of the Defects4J benchmark (b)
  used in the evaluation.}
  \label{fig:tables}
\end{figure}

In our evaluation, we use two types of oracles: \emph{representation invariants} and \emph{test assertions}. The representation invariants are taken from the Korat distribution~\cite{DBLP:conf/issta/BoyapatiKM02}, which provides 11 Java classes implementing data structures (e.g., linked lists, trees, and graphs) along with their invariants. These invariants check properties ranging from basic (e.g., no non-null values) to complex (e.g., cyclicity/acyclicity). Invariants are typically the conjunction of various properties that can be checked independently. We thus decompose each invariant into its corresponding individual properties, and consider subsets of the invariant as alternative (weaker) invariants of the same class. This yields a total of 273 distinct oracles for the Korat classes. Figure~\ref{tab:subjects-rq1-korat} summarizes the invariants used, including property counts and target invariants per class.

The test assertions in our evaluation are taken from the Defects4J benchmark~\cite{defects4j} (version 2.0.1), providing us with developer-written test assertions for real-world projects. Due to the computational cost of mutation analysis, we restrict our evaluation to test assertions from the fixed versions of the three most recent bugs in each of the 17 Defects4J projects, leading to a total of 51 versions. For each version, we consider the modified classes (and their dependencies) as target classes, and all corresponding test assertions as target oracles. Figure~\ref{tab:subjects-rq1-defects4j} summarizes the distribution of test assertions, showing, per project, the sum (across the three versions) of target classes, tests, and individual assertions.

To evaluate the correlation between state field coverage and fault detection (RQ1), we use 273 representation invariants from Korat and 17 project versions (one per project) from Defects4J, comprising 83,032 test assertions. The latter subset consists of the latest version of each project, ensuring a representative sample while reducing mutation analysis computational costs compared to analyzing all 51 versions.

For RQ2, RQ3, and RQ4, we focus on test assertions, as they better reflect real-world oracles. Since RQ2 involves mutation analysis, we reuse the same 17-project subset from RQ1. For RQ3 and RQ4, we analyze all 51 project versions to study the relationship between state field coverage and real faults, as well as the efficiency of our implementation to compute our metric. 

\subsection{Experimental Setup}

In this section we describe how we compute the metrics involved in our experiments. 

\subsubsection*{State Field Coverage} 

We compute state field coverage using the static analysis implementation from Section~\ref{sec:osc-impl}. For representation invariants, the type graph is derived from the target class (the class the invariant corresponds to). State field coverage is computed per invariant, based on its corresponding source code. For test assertions, the type graph is computed from classes modified in the bug-fixed version (as provided by Defects4J~\cite{defects4j-site}). This choice has some advantages: Defects4J mutation analysis generates mutants for these classes, facilitating our evaluation and reducing the computational cost, compared to considering \emph{all} classes of the corresponding projects. Additionally, it provides us with an unbiased criterion to select the target classes. State field coverage is computed per test, aggregating all assertions in the test. For each assertion, we inspect its code and compute covered labels, both those directly accessed and those accessed via invoked methods.

Our current implementation does not track field accesses that indirectly influence assertion parameters (e.g., via earlier statements or method calls). Detecting such cases would require more complex information flow analysis, which we leave for future work.


\subsubsection*{Mutation Analysis} 

To assess fault detection in RQ1 and RQ2, we employ mutation analysis as follows. For the representation \draftRev{invariant oracles (Korat subjects)}, mutants are generated using PIT~\cite{pit2016} \draftRev{over each of the classes, excluding the invariants themselves (invariants are the oracles, not the SUT, and thus are not mutated). We} obtained 26 mutants per class, on average. Each mutant is evaluated using a Randoop~\cite{Pacheco+2007} generated test suite (up to 100 tests per class), with each test invoking the target invariant \draftRev{as test oracle.} \draftRev{We favored the use of Randoop over EvoSuite~\cite{DBLP:conf/sigsoft/FraserA11} because of two reasons: EvoSuite's test generation is guided by mutation (among other metrics), and thus introduces a bias in the generated tests in relation to mutation analysis; also, EvoSuite aims to minimize the number of generated tests, resulting in too small test suite samples for our experiments.}

For the test assertions from Defects4J projects, mutants are generated using Major~\cite{Just+2011} (on average, $\sim$1,593 mutants per project), taking advantage of the framework's support for mutation analysis. Mutation scores are computed per-test using the test suites available with the projects (on average, $\sim$1,572 tests per project). Mutation score is computed as the percentage of mutants killed by the oracles, excluding trivial mutants triggering runtime exceptions before oracle execution.

\subsubsection*{Checked Coverage}

Since the checked coverage metric targets test assertions, we compute it for the Defects4J test assertions analyzed in RQ2. We were unable to use the original implementation~\cite{DBLP:conf/icst/SchulerZ11} due to Java version incompatibilities between the slicer used and Defects4J. Instead, we rely on a re-implementation from \cite{DBLP:conf/ast/Koitz-HristovSW22}, which is based on Slicer4J~\cite{slicer4j-site}. As discussed in the results, computation failed for some projects due to implementation errors.





\subsubsection*{Workstation}

All the experiments described in this paper were run on a workstation with a Xeon Gold 6154 CPU (3GHz), 128 GB of RAM, running Debian/GNU Linux 12. 

\subsection{Correlation with Fault Detection (RQ1)}

\begin{figure}[t]
  \centering
  \includegraphics[width=.45\textwidth]{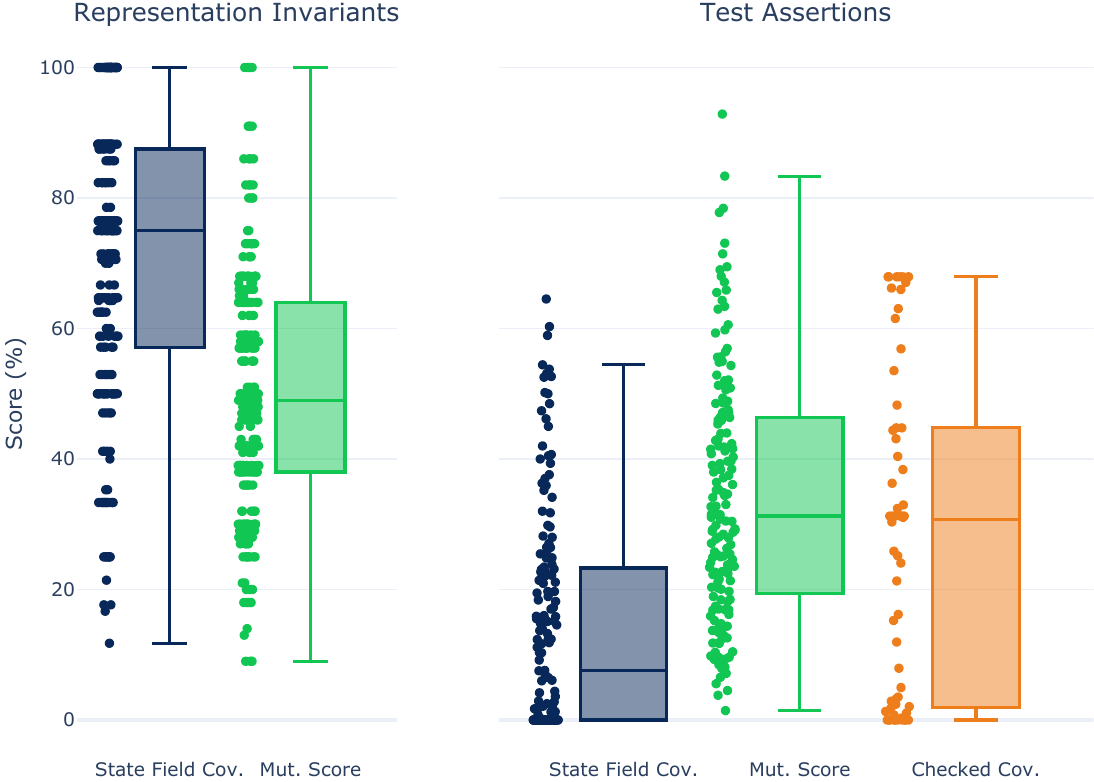}
  \caption{Distribution of the obtained State Field Coverage and Mutation Score for each type of oracle, 
  including Checked Coverage for test assertions.}
  \label{fig:box-plots-scores}
  \end{figure}

We first analyze state field coverage and mutation scores for both representation invariants and test assertions. Figure~\ref{fig:box-plots-scores} shows the results of these metrics for each oracle type, including checked coverage for test assertions. 
\draftRev{Notice that this figure shows the overall distribution of the values obtained for each metric considering all projects and whole test suites.} As mentioned earlier, for this RQ, we evaluate a subset of 17 project versions containing 83,032 test assertions; reported values on test assertions correspond to this subset.

For the 273 representation invariants analyzed, state field coverage ranges from 11.7\% to 100\% (avg. 70.3\%), while mutation scores range from 9\% to 100\% (avg. 50.2\%). The high state field coverage is expected for invariants, since these assertions explicitly check object properties, requiring access to most state fields. Indeed, 50\% of the invariants achieve over 75\% state field coverage.

The 83,032 test assertions show state field coverage ranging from 0\% to 64.5\% (avg. 14.3\%), significantly lower than invariants, as assertions typically verify method outputs corresponding to specific test inputs, rather than thoroughly inspecting object state. Over 75\% of assertions fall below 22\% state field coverage. Mutation scores range from 2.5\% to 92.8\% (avg. 34.1\%), while checked coverage spans 0\%–68\% (avg. 27.8\%).

To assess how state field coverage correlates with fault detection, we examine how increasing oracle quantity affects both state field coverage and mutation score. For representation invariants, we track these metrics as invariants grow more complex (checking additional properties). For test assertions, we evaluate them as test suites expand (adding more tests/assertions). 

\subsubsection{Correlation for Representation Invariants}


\begin{figure}[t]
  \begin{minipage}{.48\textwidth}
  \centering
  \includegraphics[width=\textwidth]{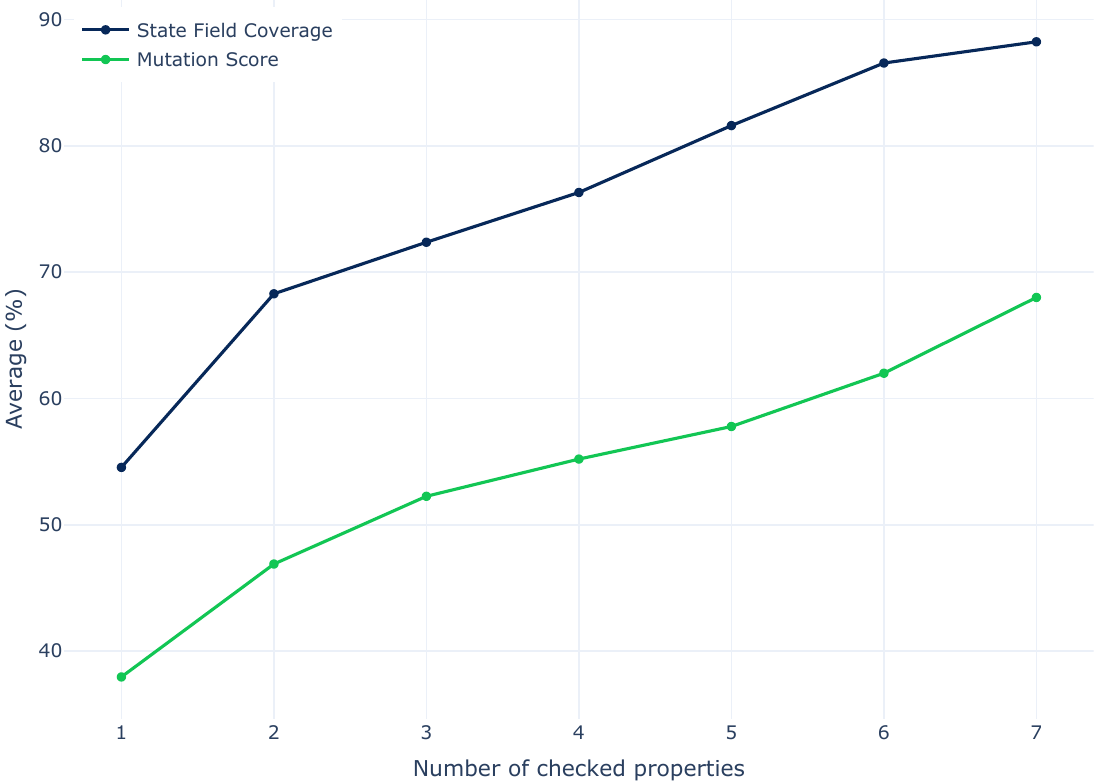}
  \caption{State Field Coverage and Mutation Score as the number of properties in representation invariants increases.}
  \label{fig:avg-by-props}
  \end{minipage}
  \hfill
  \begin{minipage}{.48\textwidth}
  \centering
  \includegraphics[width=\textwidth]{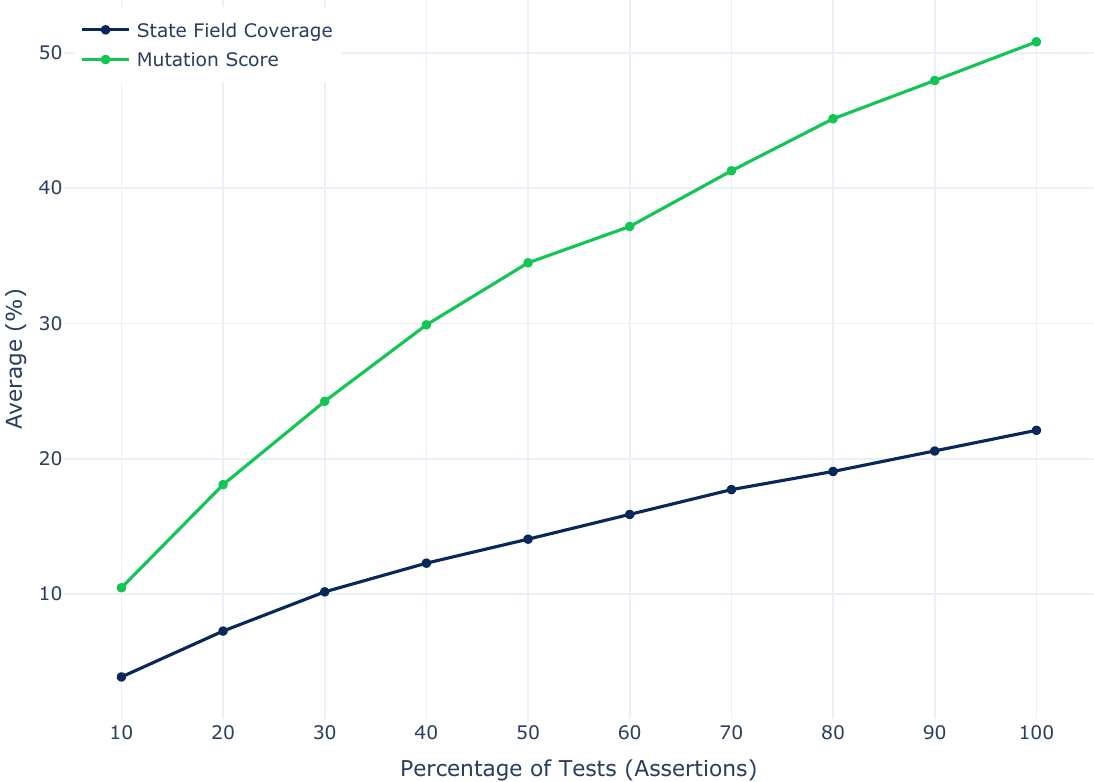}
  \caption{State Field Coverage and Mutation Score for increased percentages of selected tests and assertions.}
  \label{fig:avg-by-tests}
  \end{minipage}
  \end{figure}

Figure~\ref{fig:avg-by-props} shows how average state field coverage and mutation score increase with the number of properties checked by representation invariants. 
\draftRev{That is, we group invariants by the number 
of properties they check, and compute the 
average state field coverage and mutation score for each group.}
Both metrics exhibit similar growth trends as invariants become more complex. We quantify this relationship using Pearson correlation, which measures linear dependence between variables (ranges from -1 to 1, with values greater than zero indicating positive correlation). For our dataset, we find a coefficient of 0.54, indicating a high positive correlation between state field coverage and mutant detection.

\subsubsection{Correlation for Test Assertions}


Figure~\ref{fig:avg-by-tests} presents the relationship between test suite size (percentage of selected tests/assertions) and both state field coverage and mutation score. 
\draftRev{For each project version, we randomly select increasing percentages of tests, and compute the state field coverage and mutation score for the selected tests. 
Results are averaged across all 17 project versions.}
To account for selection randomness, each data point represents the average of 100 runs. While we analyzed state field coverage across all 17 project versions, \texttt{Collections-28} (version id 28 of the Collections project) was excluded from mutation analysis due to a Defects4J framework error that prevented score computation.

The state field coverage exhibits slower but consistent growth compared to the mutation score, as test suites grow. While their growth rates differ more markedly than with representation invariants, both metrics increase with additional tests. Calculating the Pearson correlation coefficient (excluding projects with 0\% state field coverage) yields $\sim$0.45, confirming a moderate positive correlation.

\begin{figure}[t]
  \centering
  \includegraphics[width=.45\textwidth]{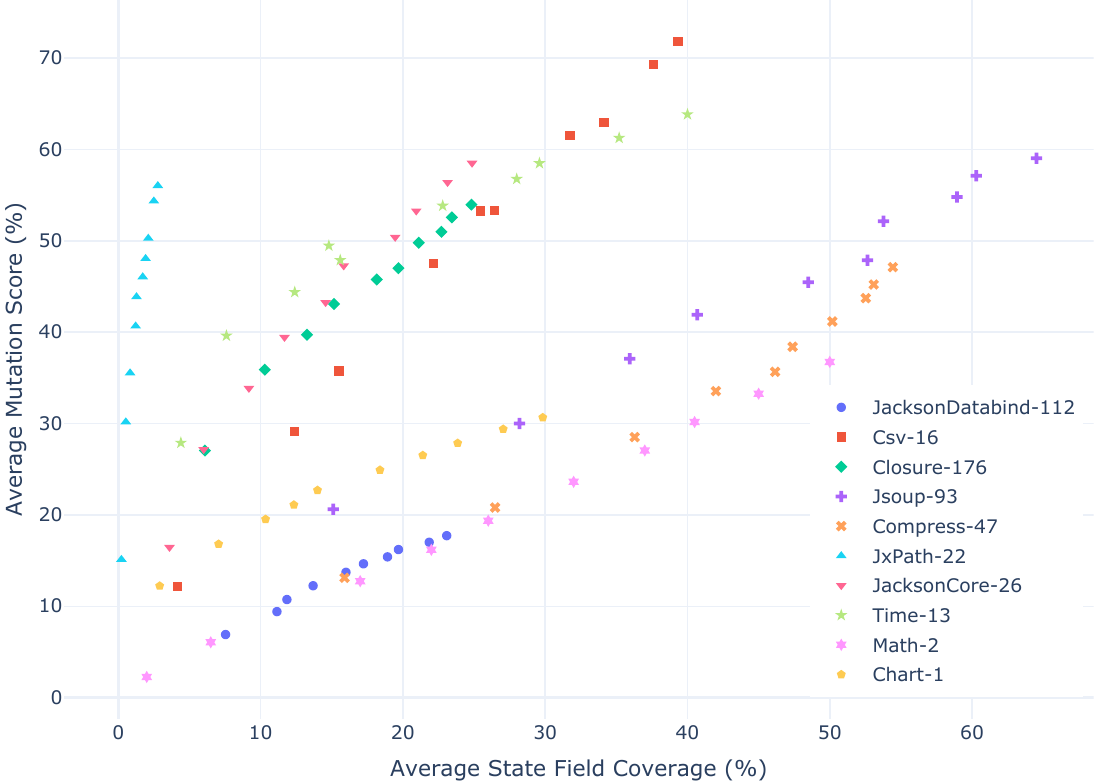}
  \caption{State Field Coverage and Mutation Score correlation 
  for each percentage of selected tests of each project version.}
  \label{fig:hc-vs-ms}
\end{figure}

For a more detailed analysis, Figure~\ref{fig:hc-vs-ms} shows per-project correlations between average state field coverage (x-axis) and average mutation score (y-axis) across test suite sizes. We excluded six projects (\texttt{Cli-40}, \texttt{Codec-18}, \texttt{Gson-16}, \texttt{JacksonXml-6}, \texttt{Lang-4}, and \texttt{Mockito-22}) where target classes were stateless (resulting in 0\% state field coverage). As discussed in Section~\ref{sec:limitations}, stateless classes lead to type graphs with no edges to cover, and thus low state field coverage does not necessarily reflect assertion limitations. 

The data shows a strong positive correlation between state field coverage and mutation score, as evidenced by the consistent trend where increased state field coverage improves fault detection. This relationship is particularly robust, with Pearson coefficients exceeding 0.96 for most projects. The sole exception is \texttt{JxPath-22}, where maximum coverage plateaus at 2.78\% due to tests overall covering a small number of labels. In this case, additional tests cannot significantly improve state field coverage, allowing mutation scores to increase independently. These results demonstrate that state field coverage effectively predicts oracle quality for both representation invariants and test assertions, showing consistent correlation with the established mutation score metric. 

\subsection{Oracle Improvement (RQ2)}

To assess how state field coverage can guide oracle improvement, we conduct an experiment using test assertions. We simulate extending an initial single-test suite by incrementally adding tests \draftRev{whose oracles} maximize uncovered state field labels, \draftRev{comparing} the resulting mutation score against \draftRev{that of} random test selection (averaged over 10 iterations to account for randomness). Since this experiment also involves mutation analysis, we use the same 17 project versions from RQ1. \draftRev{When possible, we} simulate \draftRev{oracle guided} test suite extension based on checked coverage, \draftRev{reporting} the results for cases where \draftRev{this metric was successfully computed} (projects \texttt{Csv-16}, \texttt{Jsoup-93}, and \texttt{Time-13} out of all the projects in this experiment). \draftRev{Note that \texttt{Collections-28} had to be excluded due to a Defects4J error, and other six projects are disregarded due to the corresponding modified classes being stateless}. \draftRev{Additionally, we incorporate as a reference a test suite extension strategy based on optimizing statement coverage. Notice that this strategy chooses as new tests to extend the suite those that maximize covering previously uncovered statements, without considering whether these newly covered statements are called from oracles or not. Thus, it is not an oracle guided strategy, but serves as a reference with traditional test prioritization strategies.}

\begin{figure*}[t]
  \centering
  \includegraphics[width=0.95\textwidth]{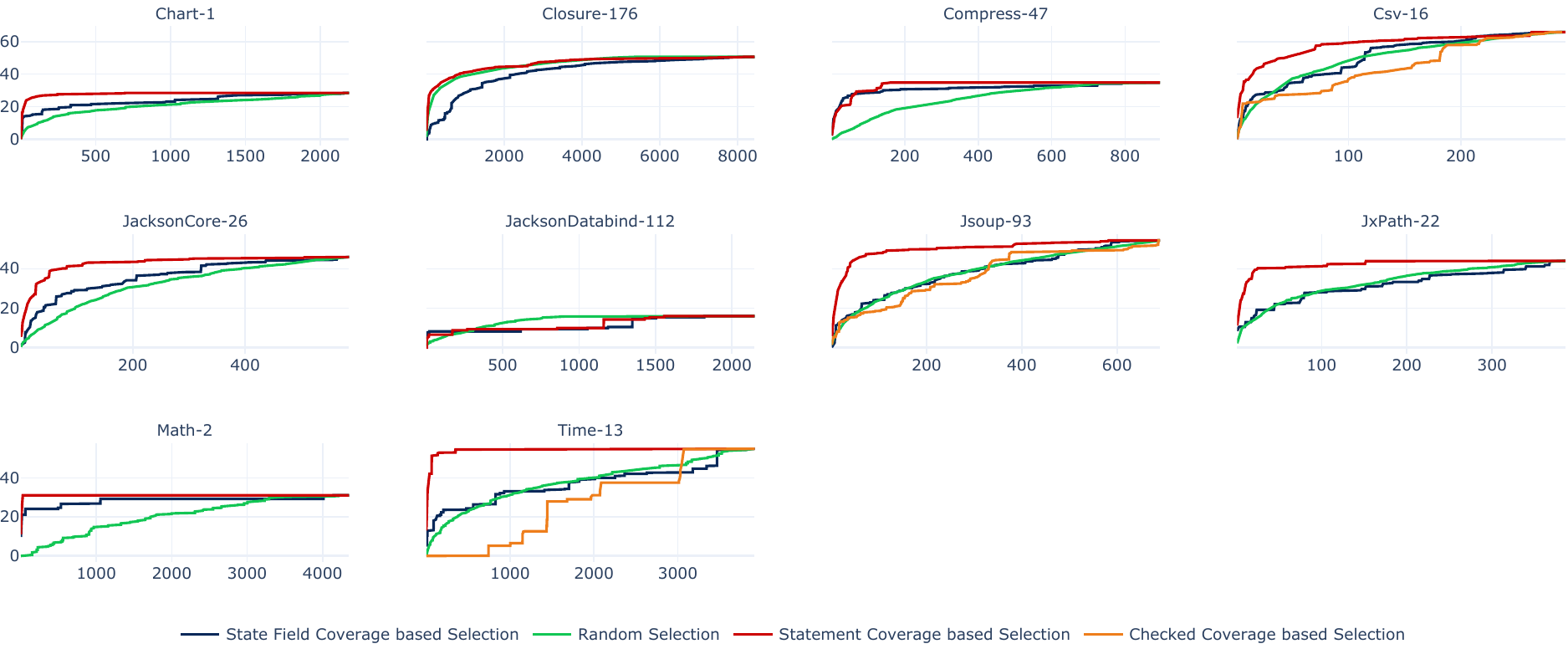}
  \caption{Mutation score achieved by selecting tests that increasingly improve State Field Coverage, compared to selecting tests randomly. A selection based on improving Checked Coverage is also included when possible. The x axis represents the number of selected tests, and the y axis represents the mutation score achieved by the tests. Random selection is iterated 10 times.}
  \label{fig:hc-vs-random}
\end{figure*}

Figure~\ref{fig:hc-vs-random} shows the experimental results across all projects, plotting mutation score (y-axis) against number of selected tests (x-axis). 
The graph compares four test selection strategies: 
state field coverage maximization (blue line), 
random selection (green line), 
checked coverage maximization (orange line, where available), 
\draftRev{and statement coverage maximization (red line).}
We analyze each strategy's effectiveness below. 

\subsubsection{Statement Coverage-based Selection}

\draftRev{Increasingly selecting tests that maximize statement coverage leads to higher mutation scores compared to the other strategies in most of the project versions. This is expected, as improving statement reachability improves mutant detection, and this approach does not focus on the oracles, but on the statements covered by the unit tests as a whole. The purpose of including this strategy is to provide a reference point for comparison, as is not directly related to oracle quality. As we discuss below, state field coverage-based selection is in general superior to the random and checked coverage-based selection approaches.}

\subsubsection{State Field Coverage-based Selection}


Several project versions demonstrate significantly higher mutation scores when extending test suites via state field coverage maximization versus random selection. In \texttt{Chart-1}, \texttt{Math-2}, \texttt{Compress-47}, and \texttt{JacksonCore-26}, this strategy detects up to 20\% more faults initially. The mutation score improvement remains substantial even at scale, e.g., for \texttt{Chart-1}, \texttt{Compress-47}, and \texttt{JacksonCore-26} with 300 tests, and for \texttt{Math-2} with 2,000 tests.

\begin{figure}[t]
  \centering
  \scriptsize
\begin{lstlisting}[language=Java,linewidth={\linewidth},frame=tb]
@Test
public void testMoments() {
  final double tol = 1e-9;
  HypergeometricDistribution dist;
  
  dist = new HypergeometricDistribution(1500, 40, 100);
  Assert.assertEquals(dist.getNumericalMean(), 40d * 100d / 1500d, tol);
  Assert.assertEquals(dist.getNumericalVariance(), ( 100d * 40d * (1500d - 100d) * (1500d - 40d) ) / ( (1500d * 1500d * 1499d) ), tol);
  
  dist = new HypergeometricDistribution(3000, 55, 200);
  Assert.assertEquals(dist.getNumericalMean(), 55d * 200d / 3000d, tol);
  Assert.assertEquals(dist.getNumericalVariance(), ( 200d * 55d * (3000d - 200d) * (3000d - 55d) ) / ( (3000d * 3000d * 2999d) ), tol);
}
\end{lstlisting}
  \caption{Test from \texttt{Math-2} with a 50\% state field coverage and 9.7\% mutation score.}
  \label{fig:math-testMoments-test}
  \vspace{-1em}
\end{figure}

Figure~\ref{fig:math-testMoments-test} illustrates a representative test from \texttt{Math-2} that was prioritized by our state field coverage heuristic. This single test achieves a 9.7\% mutation score (30\% of the suite's total 31.08\%) while covering 50\% of target class labels with just four assertions. Similar patterns emerge in other projects: in \texttt{Chart-1}, the first test yields $\sim$3\% mutation score, increasing to $\sim$10\% with 10 tests and $\sim$20\% with 100 tests, nearing the maximum $\sim$30\%. For \texttt{Compress-47}, the first ten tests achieve $\sim$16.3\% mutation score, a notable efficiency given the suite's 800+ tests and maximum \draftRev{$\sim$}34.8\% score.


In several project versions (\texttt{JxPath-22}, \texttt{JacksonDatabind-112}, \texttt{Time-13}, \texttt{Csv-16}, and \texttt{Jsoup-93}), our state field coverage strategy yields mutation scores comparable to random selection. These projects generally exhibit low state field coverage (e.g., below 25\% for \texttt{JxPath-22} and \texttt{JacksonDatabind-112}), suggesting that our metric provides limited benefit when state field coverage is low. However, for projects like \texttt{JacksonDatabind-112} and \texttt{Csv-16}, our strategy achieves better initial mutation scores.
The \texttt{Closure-176} project is unique in showing significantly worse performance with our strategy compared to random selection. This aligns with its very low state field coverage ($\sim$25\%), among the poorest of all projects.


These results demonstrate that state field coverage can effectively guide oracle improvement, particularly when high state field coverage is achievable. Furthermore, our metric's ability to identify uncovered state portions directly supports targeted assertion generation for improved fault detection.

\subsubsection{Checked Coverage-based Selection}

Figure~\ref{fig:hc-vs-random} includes checked-coverage based selection results for \texttt{Time-13}, \texttt{Csv-16}, and \texttt{Jsoup-93}, the only projects where both checked coverage and state field coverage could be computed. The analysis shows that for \texttt{Csv-16} and \texttt{Jsoup-93}, checked coverage yields comparable mutation scores to state field coverage in early test selection phases, while \texttt{Time-13} shows inferior performance for checked coverage compared to the other strategies. 
While state field coverage generally outperforms checked coverage in our experiments, more extensive studies are required to validate this observed trend, and better characterize the relationship between these metrics. 

\begin{table*}[t!]
\scriptsize
\setlength{\tabcolsep}{4.5pt}
\renewcommand{\arraystretch}{1.2}
\centering
\caption{Weighted Average of the Percentage of Faults Detected (APFD) by
the State Field Coverage-based Selection (SFC) and Random Selection (Random).
For each target project, we report the APFD values achieved by each strategy
considering a percentage of selected tests. 
Green SFC cells indicate that SFC outperforms random selection, red cells otherwise.
}
\begin{tabular}{llc|c|c|c|c|c|c|c|c|c}
\toprule
\multirow{2}{*}{\textbf{Subject}} & 
\multirow{2}{*}{\textbf{Technique}} & 
\multicolumn{10}{c}{\textbf{APFD by Test suite percentages}} \\
& & \textbf{10} & \textbf{20} & \textbf{30} & \textbf{40} & \textbf{50} & \textbf{60} & \textbf{70} & \textbf{80} & \textbf{90} & \textbf{100} \\
\midrule
Chart-1 & SFC & \comparecell{67.75}{81.98} & \comparecell{68.58}{67.37} & \comparecell{76.95}{72.49} & \comparecell{79.79}{78.88} & \comparecell{83.3}{82.55} & \comparecell{37.46}{83.73} & \comparecell{46.25}{83.16} & \comparecell{52.96}{44.72} & \comparecell{57.8}{50.25} & \comparecell{61.26}{55.09} \\
& Random & 81.98 & 67.37 & 72.49 & 78.88 & 82.55 & 83.73 & 83.16 & 44.72 & 50.25 & 55.09 \\ \hline
Closure-176 & SFC & \comparecell{65.36}{63.82} & \comparecell{75.55}{76.2} & \comparecell{79.31}{81.06} & \comparecell{81.81}{81.74} & \comparecell{83.47}{84.02} & \comparecell{78.45}{84.43} & \comparecell{78.91}{85.49} & \comparecell{79.96}{85.04} & \comparecell{81.14}{85.25} & \comparecell{80.92}{86.08} \\
& Random & 63.82 & 76.2 & 81.06 & 81.74 & 84.02 & 84.43 & 85.49 & 85.04 & 85.25 & 86.08 \\ \hline
Compress-47 & SFC & \comparecell{62.16}{31.25} & \comparecell{60.71}{40.86} & \comparecell{63.41}{57.73} & \comparecell{67.53}{60.34} & \comparecell{67.69}{51.54} & \comparecell{70.43}{54.18} & \comparecell{74.91}{54.51} & \comparecell{75.93}{54.22} & \comparecell{77.48}{58.35} & \comparecell{78.27}{57.75} \\
& Random & 31.25 & 40.86 & 57.73 & 60.34 & 51.54 & 54.18 & 54.51 & 54.22 & 58.35 & 57.75 \\ \hline
Csv-16 & SFC & \comparecell{74.84}{52.79} & \comparecell{77.61}{65.9} & \comparecell{73.82}{71.51} & \comparecell{73.57}{72.39} & \comparecell{66.76}{75.07} & \comparecell{68.59}{75.1} & \comparecell{71.83}{76.11} & \comparecell{72.5}{76.33} & \comparecell{73.76}{76.21} & \comparecell{75.11}{77.48} \\
& Random & 52.79 & 65.9 & 71.51 & 72.39 & 75.07 & 75.1 & 76.11 & 76.33 & 76.21 & 77.48 \\ \hline
JacksonCore-26 & SFC & \comparecell{64.7}{52.75} & \comparecell{76.88}{61.03} & \comparecell{77.13}{62.47} & \comparecell{74.3}{65.74} & \comparecell{74.61}{67.07} & \comparecell{77.09}{65.35} & \comparecell{77.04}{66.46} & \comparecell{77.97}{67.44} & \comparecell{79.89}{70.24} & \comparecell{79.74}{71.84} \\
& Random & 52.75 & 61.03 & 62.47 & 65.74 & 67.07 & 65.35 & 66.46 & 67.44 & 70.24 & 71.84 \\ \hline
JacksonDatabind-112 & SFC & \comparecell{98.74}{67.7} & \comparecell{94.7}{60.6} & \comparecell{96.46}{61.41} & \comparecell{66.67}{56.82} & \comparecell{70.45}{58.07} & \comparecell{75.37}{60.6} & \comparecell{73.66}{57.28} & \comparecell{76.94}{62.66} & \comparecell{74.61}{64.74} & \comparecell{77.16}{68.29} \\
& Random & 67.7 & 60.6 & 61.41 & 56.82 & 58.07 & 60.6 & 57.28 & 62.66 & 64.74 & 68.29 \\ \hline
Jsoup-93 & SFC & \comparecell{69.96}{51.54} & \comparecell{72.65}{59.41} & \comparecell{64.08}{70.73} & \comparecell{63.24}{70.96} & \comparecell{66.97}{70.2} & \comparecell{70.06}{70.96} & \comparecell{69.55}{71.88} & \comparecell{71.84}{72.15} & \comparecell{72.47}{74.33} & \comparecell{73.97}{74.68} \\
& Random & 51.54 & 59.41 & 70.73 & 70.96 & 70.2 & 70.96 & 71.88 & 72.15 & 74.33 & 74.68 \\ \hline
JxPath-22 & SFC & \comparecell{66.45}{56.53} & \comparecell{80.81}{75.39} & \comparecell{83.44}{76.21} & \comparecell{87.07}{70.77} & \comparecell{73.54}{73.34} & \comparecell{70.6}{76.69} & \comparecell{74.86}{78.97} & \comparecell{73.83}{72.31} & \comparecell{70.13}{74.3} & \comparecell{69.01}{76.9} \\
& Random & 56.53 & 75.39 & 76.21 & 70.77 & 73.34 & 76.69 & 78.97 & 72.31 & 74.3 & 76.9 \\ \hline
Math-2 & SFC & \comparecell{94.51}{7.01} & \comparecell{97.25}{46.06} & \comparecell{98.17}{58.63} & \comparecell{88.21}{48.69} & \comparecell{90.02}{44.11} & \comparecell{91.69}{46.69} & \comparecell{83.66}{54.31} & \comparecell{85.71}{60.02} & \comparecell{85.72}{63.27} & \comparecell{87.14}{65.45} \\
& Random & 7.01 & 46.06 & 58.63 & 48.69 & 44.11 & 46.69 & 54.31 & 60.02 & 63.27 & 65.45 \\ \hline
Time-13 & SFC & \comparecell{38.42}{57.03} & \comparecell{69}{68.55} & \comparecell{79.2}{77.49} & \comparecell{76.79}{83.22} & \comparecell{80.63}{86.63} & \comparecell{83.8}{83.7} & \comparecell{81.09}{86.07} & \comparecell{83.41}{75.46} & \comparecell{85.28}{57.42} & \comparecell{86.78}{58.54} \\
& Random & 57.03 & 68.55 & 77.49 & 83.22 & 86.63 & 83.7 & 86.07 & 75.46 & 57.42 & 58.54 \\ \midrule
\multicolumn{2}{c}{Times SFC $>$ Random:} & \textbf{8} & \textbf{9} & \textbf{8} & \textbf{8} & \textbf{6} & \textbf{5} & \textbf{4} & \textbf{7} & \textbf{6} & \textbf{6} \\
\bottomrule
\end{tabular}
\label{tab:apfd}
\end{table*}

\draftRev{Finally, we analyze whether the effect of state field coverage (SFC) based selection is due to indirectly improving code coverage, or not. We build test suites where all tests have very similar statement coverage, selecting the maximum subset of tests where the difference in statement coverage among any two tests is at most 10\%. Then, we perform SFC-based and random selections from these subsets. To compare these strategies, we compute the progression of the APFD metric~\cite{apfd-paper}, which measures a weighted average of the percentage of faults detected, for increasingly larger subsets (10\%, 20\%, and so on), of the suite with similar statement coverage. The results of this experiment are shown in Table~\ref{tab:apfd}. 

Notably, for the first percentages of selected tests (10\% to 40\%), SFC selection considerably outperforms random selection in most of the project versions, obtaining a higher APFD in at least 8 out of 10 project versions. This indicates that, for the same level of code coverage, selecting tests that improve oracles according to state field coverage leads to better fault detection than selecting tests randomly. As the percentage of selected tests increases, the advantage of state field coverage-based selection diminishes, but still outperforms random selection in most projects. These results show that our metric provides benefits beyond code coverage, and can effectively guide test selection for improved fault detection. }

\draftRev{It is worth remarking that, in some cases, the progression of APFD for greater suite subsets can decrease. For instance, in \texttt{Chart-1}, the APFD progression for SFC drops from 83.3 to 37.46, when progressing from 50\% to 60\%. While this may be counter intuitive, it is indeed possible. It is an effect due to new tests being incorporated, that kill new mutants, but do so with the latest tests in the selection order, thus causing the average percentage of faults detected to drop. This issue is also observed for random selection, where the APFD drops from 83.14 to 44.72 when moving from 70\% to 80\%.}

\subsection{Real Fault Detection (RQ3)}

To evaluate our metric's impact on real fault detection, we conduct the following experiment using Defects4J buggy versions with failing tests. We measure how many test executions are needed to trigger a bug under two strategies: \emph{state field coverage-based execution}, where tests are ordered to maximize state field coverage growth (as in RQ2), and \emph{random execution}, where tests are selected randomly (averaged over 10 runs). From the initial 51 project versions, 38 (75\%) have non-empty type graphs (enabling state field coverage computation), 27 of which (71\% of 38) yield a positive state field coverage. We focus our analysis on these 27 versions, since as discussed in Section~\ref{sec:limitations}, our metric cannot be computed for stateless classes. 

\begin{table}[t]
  \scriptsize
  \centering
  \caption{Test cases needed to first trigger Defects4J bugs, when selecting tests based on State Field Coverage, and on Random selection.}
  \begin{tabular}{lrr|rr}
  \toprule
  \multirow{2}{*}{\textbf{Project}} & 
  \multirow{2}{*}{\textbf{Bug ID}} & 
  \multirow{2}{*}{\textbf{\#Tests}} & \multicolumn{2}{c}{\textbf{Tests needed to trigger the bug}} \\ 
  & & & \textbf{State Field Cov.} & \textbf{Random} \\
  \midrule
  Chart & 1 & 2,193 & 4 & 781.5 \\
  Chart & 3 & 2,187 & 16 & 1,423.9 \\
Cli & 38 & 317 & 56 & 97.8 \\
Closure & 174 & 8,308 & 1,337 & 2,081.3 \\
Closure & 175 & 8,410 & 231 & 1,098.4 \\
Closure & 176 & 8,432 & 5,895 & 4,579.4 \\
Collections & 26 & 2,720 & 171 & 1,364.3 \\
Compress & 46 & 829 & 2 & 506.9 \\
Compress & 47 & 895 & 32 & 461.6 \\
Csv & 14 & 257 & 97 & 33.5 \\
Csv & 15 & 290 & 212 & 190 \\
Csv & 16 & 293 & 88 & 114.6 \\
JacksonCore & 26 & 585 & 275 & 333.7 \\
JacksonCore & 25 & 573 & 57 & 285.8 \\
JacksonCore & 24 & 580 & 109 & 58.6 \\
JacksonDatabind & 111 & 2,146 & 621 & 946.4 \\
JacksonDatabind & 112 & 2,148 & 2,145 & 973.5 \\
Jsoup & 92 & 689 & 317 & 135.8 \\
Jsoup & 93 & 690 & 309 & 377.5 \\
JxPath & 21 & 384 & 76 & 144.6 \\
JxPath & 22 & 386 & 347 & 168.7 \\
Math & 1 & 4,378 & 185 & 1,252.2 \\
Math & 2 & 4,350 & 3,000 & 2,463.4 \\
Mockito & 1 & 1,370 & 49 & 60.2 \\
Time & 1 & 4,041 & 2,258 & 1,696.8 \\
Time & 2 & 4,041 & 1,031 & 1,923.2 \\
Time & 13 & 3,916 & 3,705 & 2,035.2 \\
  \midrule
  \multirow{2}{*}{\textbf{Summary}} & 
  \multicolumn{2}{r}{Times better:} & \textbf{17} & 10 \\
  & \multicolumn{2}{r}{Average improvement:} & \textbf{34.7x} & 1.8x \\
  \bottomrule
  \end{tabular}
  \label{tab:defects4j-real-bugs}
  \vspace{-1em}
\end{table}

Table~\ref{tab:defects4j-real-bugs} presents the results comparing test selection strategies. For each project and bug id, we show the total tests available, and the number of tests needed to first trigger the bug using state field coverage ordering, and random ordering. The state field coverage ordering outperformed random selection in 17/27 cases (63\%), requiring 34.7× fewer tests on average. In the remaining 10 cases, random selection performed modestly better (1.8× fewer tests). This demonstrates our metric's potential to significantly improve fault detection efficiency by prioritizing tests more likely to reveal bugs.

\subsection{Efficiency (RQ4)}

Finally, we evaluate the execution time required to (statically) compute state field coverage, and compare it with mutation analysis and checked coverage. Our evaluation involves test assertions from the Defects4J projects. For the 17 projects analyzed, state field coverage computation required an average of 5,164.0 seconds ($\sim$1.4 hours) per project, totaling 82,624.2 seconds ($\sim$22 hours) for all 83,032 test assertions ($\sim$6 seconds per test). This contrasts with mutation analysis, which averaged 72,902.9 seconds ($\sim$20 hours) per project and exceeded 300 hours ($\sim$12.5 days) total due to its inherently dynamic nature. Similarly, checked coverage (measured for the 6 projects for which the checked coverage tool could be run) averaged 20,327.3 seconds ($\sim$5.6 hours) per project, totaling 121,964.1 seconds, with its dynamic slicing process contributing to the higher computation time. These results demonstrate that static field coverage provides rapid, practical feedback on oracle quality, enabling efficient preliminary assessment before committing to more computationally expensive analyses like mutation analysis.

\section{Limitations}
\label{sec:limitations}


Our metric requires the target class to be stateful, containing fields accessed by the class methods. Stateless classes (e.g., \texttt{Cli-40}, \texttt{Codec-18}, \texttt{Gson-16}, and \texttt{Mockito-22}) with no fields and only static methods yield empty type graphs and consequently 0\% state field coverage. However, this limitation affects only a minority of cases: 38 of the 51 analyzed project versions (75\%) contained stateful classes with at least one field, demonstrating our technique's broad applicability. Also, our current metric definition considers only the target class's fields and their direct and indirect dependencies, but excluding types returned by methods. Consequently, assertions that only verify method return values yield 0\% state field coverage, limiting the metric's ability to assess such oracles. 
We plan to extend the metric to include return type fields, which would both address this limitation and resolve the stateless class issue, by analyzing returned objects' state. This enhancement, part of our future work, would further broaden the metric's applicability to more oracle types and target classes.

\draftRev{Another limitation of our approach is the potential infeasibility of state fields as test requirements. Unlike mutation testing, where infeasible requirements (e.g., equivalent mutants) are inherent, state field coverage infeasibility would correspond to fields that are unreachable from test oracles. Such infeasibility highlights limitations in the testability of the system under test (e.g., poorly exposed state), rather than a flaw in the metric itself. This behavior is analogous to unreachable code in statement coverage metrics and reflects useful diagnostic information rather than a weakness.}

\section{Threats To Validity}

A threat to external validity stems from our use of a Defects4J subset rather than all available projects, limited by mutation analysis costs. However, we included a representative variety of 215 target classes. Implementation issues with checked coverage also constrained our comparison, despite best efforts to use available implementations. 

For internal validity, potential threats include: \emph{(1)} our static implementation may conservatively include field access paths not executed at runtime, potentially inflating metric values; \emph{(2)} the observed correlation between state field coverage and mutation score might indirectly stem from improved code coverage; and \emph{(3)} stochastic variations in our experiments could inadvertently bias results. To address these threats, we implemented the following mitigation strategies. For \emph{(1)}, we developed a dynamic variant of state field coverage and compared it against our static metric. The two agreed exactly in 40\% of cases and differed by $<$10\% in the remaining 60\%, demonstrating strong alignment. For \emph{(2)}, we conducted controlled experiments by clustering tests with identical statement coverage, then rerunning the state field coverage and random selection strategies within these clusters. The results reaffirmed our original correlation findings, isolating the effect of state field coverage. For \emph{(3)}, we repeated randomized trials and manually verified outcomes to minimize chance effects. Summaries of the additional experiments can be found as part of our replication package. 

\section{Conclusion and Future Work}
    
Testing effectiveness ultimately resorts on oracle quality. Existing
techniques, such as mutation analysis, checked coverage, and oracle
deficiency, have advanced oracle assessment but are computationally costly and often provide indirect, hard-to-act-on feedback.
We introduced \emph{state field coverage}, a novel metric for assessing oracle quality based on the premise that oracles referencing more of the SUT's state definition are more effective at detecting faults. Our static approach to compute this metric addresses a key limitation of dynamic techniques, by avoiding their computational costs while maintaining strong correlation with fault detection (as validated through mutation analysis). Unlike existing methods, our metric directly identifies uncovered state elements, providing developers with actionable insights for oracle improvement.
Future work includes extending SFC to additional oracle forms (e.g., properties in property-based testing~\cite{DBLP:conf/icfp/ClaessenH00}), broadening the empirical evaluation, and integrating SFC as a fitness signal in evolutionary test generation (e.g., EvoSuite~\cite{DBLP:conf/sigsoft/FraserA11}) to favor tests whose assertions more thoroughly predicate on program state.

\section{Data Availability}

\draftRev{Our current state field coverage implementation as well as
  the scripts and data required to reproduce our experiments are
  publicly available in our replication package~\cite{osc-site}.}

\section*{Acknowledgments}

This work is supported by the Ramón y Cajal fellowship
RYC2020-030800-I, by the Spanish Government through grants
TED2021-132464B-I00 (PRODIGY), PID2022-142290OB-I00 (ESPADA), and
CEX2024-001471-M/funded by MICIU/AEI/10.13039/501100011033 and by the
Comunidad de Madrid as part of the ASCEND project co-funded by FEDER
Funds of the European Union.

\bibliographystyle{plain}
\bibliography{references}

\end{document}